\documentstyle[art12]{article}
\begin{document}
\title{Conservation laws for 
linear equations on quantum Minkowski 
spaces\footnote{Research partially supported
by KBN grant 2P 302 087 06}}
\author{M. Klimek \\
Institute of Mathematics and Computer Science \\ Technical University
of Cz\c{e}stochowa, \\ 
ul.D\c{a}browskiego 73, 42-200 Cz\c{e}stochowa, Poland \\
E-mail: klimek@matinf.pcz.czest.pl}
\date{}
\maketitle
\begin{abstract}
The  general, linear
equations with constant coefficients on quantum Minkowski spaces 
are considered and the explicit formulae for their conserved currents
are given. The proposed procedure can be simplified
for $ * $-invariant equations.\\
The derived method is then  applied to Klein-Gordon, Dirac and wave equations
on different classes of Minkowski spaces. In the examples
also symmetry operators for these equations are obtained. They include
quantum deformations of classical symmetry operators as well as an additional
operator connected with deformation of the Leibnitz rule in non-commutative
differential calculus.
\end{abstract}
\section{Introduction}
The investigation of conservation laws 
and invariants of motion for given action 
or equation of motion is an important part  of classical mechanics and 
field theory. The general problem is solved by Noether theorem in which
symmetry of the action yields conserved currents and integrals
of motion. When we restrict our study to linear equations there 
is known method
of Takahashi and Umezawa \cite{a0} which allows us to construct
invariants for classical field-theoretic models.
We have shown previously that it can be extended to discrete and mixed
models on commutative spaces \cite{a,b,c}. 
The equations of this kind appear also in realizations of generators
and Casimir operators of quantum algebras
on commutative spaces, namely for $ \kappa $-deformed
algebras  \cite{v,w,x}.\\
The aim of this paper is to extend this method to linear equations on 
quantum Minkowski spaces. We have choosen  to work within the framework
of class of Minkowski spaces endowed with the action of 
quantum Poincar\'{e} groups, which were
 introduced by Woronowicz and Podle\'{s}
in \cite{p,r,s} and use the differential calculus developed by Podle\'{s}
in \cite{o}. Let us however notice that once the explicit formula
of Leibnitz rule for exterior derivatives  for other quantum spaces
and differential calculi is given, one can extend the proposed procedure. 
The example of this type is the braided differential calculus
on q-Minkowski space introduced in 
the work by Ogievetsky {\em{et al.}} \cite{ab}
and formally developed by Majid  \cite{m,n}. \\
Let us sketch briefly the steps of the procedure:
\begin{itemize}
\item{in derivation of the conserved currents
Leibnitz rule is used. In case of non-commutative differential calculus
it is deformed similarly  as in the discrete calculus \cite{a,c}. We introduce
modified Leibnitz rule}
\item{the special operator $ \Gamma $ 
is built from derivatives acting on the right- 
and left-hand side. In quantum case we must use derivatives and their
conjugations}
\item{for hermitian equation operators the hermitian currents can be derived.
As the explicit form of scalar product on quantum Minkowski space
is not known we shall use throughout the paper the norion of $ * $ -invariant
equation operator for which the construction can be simplified}
\item{to obtain different solutions for given equation one needs its symmetry 
operators. Their algebras are well known 
in classical models; we show in examples
the symmetry operators for Klein-Gordon, Dirac and wave equations without
discussion of their algebraic properties. It was however shown
in example of Klein-Gordon equation on quantum Minkowski space with $ Z=0 $
that the algebra closes \cite{d}.}
\end{itemize}
\section{Modification of Leibnitz rule in differential calculi 
on quantum Minkowski spaces}
In the paper we shall work within the framework of differential calculi
on quantum Minkowski spaces, in general case  introduced and 
investigated in \cite{o}. Let us remind the fundamental rules of commutation
for partial derivatives and variables:
\begin{eqnarray}
 & \partial^{j} (x^{i})=g^{ji} & \\
 & \partial^{l} \partial^{k}= 
 R^{lk}_{\;\;\;\;ij} \partial^{i} \partial^{j} &  \label{comp} \\
 & \partial^{i} x^{k}=
 g^{ik}+(R^{ik}_{\;\;\;\; ab}x^{a} 
 -(RZ)^{ik}_{\;\;\;\; b}) \partial^{b} &  \label{com} \\
 & (R-1)^{ij}_{\;\;\;\; kl}[x^{k} x^{l} 
 -Z^{kl}_{\;\;\;\; s}x^{s}+T^{kl}]=0 & \label{comx} 
\end{eqnarray} 
where R-matrix fulfills quantum Yang-Baxter 
equation $ R_{23}R_{12}R_{23}=R_{12}R_{23}R_{12} $
and the condition $ R^{2}=1 $. The metric 
tensor $ g $ appearing in the above formulae
is R-symmetric, that means $ Rg=g $. \\
The functions on quantum Minkowski 
space are understood as formal power series
of monomials of variables. For the product of two such arbitrary functions 
the partial derivatives obey the following Leibnitz rule \cite{o}:
\begin{equation}
\partial^{i} fg=(\partial^{i} f)g+(\zeta^{i}_{j} f)\partial^{j}g
\label{leibder}
\end{equation}
with the transformation operator $ \zeta $ fulfilling the equality:
\begin{equation}
\zeta^{i}_{j}(fg)=(\zeta^{i}_{m}f)(\zeta^{m}_{j} g)
\label{mult}
\end{equation}
and connected with the operator $ \rho $ from \cite{o} via the equation:
\begin{equation}
\zeta^{i}_{j} =g^{ia} \rho^{b}_{a} g_{bj}
\end{equation}
Formula (\ref{com}) yields the e
xplicit form of transformation operator acting on the monomial of 
the first order:
\begin{equation}
\zeta^{i}_{j} x^{k}=R^{ik}_{\;\;\;\; aj}x^{a} -(RZ)^{ik}_{\;\;\;\; j}
\label{transf}
\end{equation}
and using  (\ref{mult}) 
can be easily extended to arbitrary function on quantum Minkowski space. \\
It is interesting point 
that (\ref{comx}) implies analogous Leibnitz rule to be valid also
for variables, namely:
\begin{equation}
x^{i}(fg)=(x^{i}f)g=(x^{i}f)_{act}g+(\tilde{\zeta}^{i}_{j}f)x^{j}g
\label{leibx}
\end{equation}
where the action of variable on monomial is due to the selfineraction of 
variables, is determined by properties of quantum Minkowski space (\ref{comx}) 
and looks as follows:
\begin{equation}
(x^{i}[k_{1},...,k_{n}])_{act}
=\sum_{l=1}^{n}(\tilde{\zeta}^{i}_{j}[k_{1},...k_{l-1}])
v^{jk_{l}} [k_{l+1},...,k_{n}]
\label{xact}
\end{equation}
The initial and last term are of the form:
\begin{equation}
l=1 \hspace{1cm} v^{ik_{1}}[k_{2},...,k_{n}] \hspace{2cm}
l=n \hspace{1cm} \tilde{\zeta}^{i}_{j} [k_{1},...,k_{n-1}]v^{jk_{n}}
\end{equation}
and appearing in the formulae 
(\ref{leibx},\ref{xact}) transformation operator $ \tilde{\zeta} $
is defined for arbitrary function 
by its action on $ x^{k}  $ and multiplicity property:
\begin{eqnarray}
 & \tilde{\zeta}^{i}_{j} x^{k}=
 R^{ik}_{\;\;\;\; a j}x^{a}+(Z-RZ)^{ik}_{\;\;\; j}  & \\
 & \tilde{\zeta}^{i}_{j} (fg)
 =(\tilde{\zeta}^{i}_{m} f)(\tilde{\zeta}^{m}_{j} g)
\end{eqnarray}
In the case where the selfinteraction 
of variables determined by tensor $ v=RT-T $
vanishes,
the Leibnitz rule for variables reduces to the formula:
\begin{equation}
x^{i}(fg)=(x^{i}f)g=(\tilde{\zeta}^{i}_{j}f)x^{j}g
\end{equation}
Finally we can also write both 
Leibnitz rules (\ref{leibder},\ref{leibx}) in the vector form:
\begin{eqnarray}
& \vec{\partial} fg=(\vec{\partial} f)g+(\zeta f)\vec{\partial} g & \\
& \vec{x} fg=(\vec{x}f)g=(\vec{x}f)_{act} g+(\tilde{\zeta}f)\vec{x} g & 
\end{eqnarray}
In the next sections we shall consider linear  equations of motion and derive
for them the conservation laws. We have solved similar problem in  
discrete models on commutative spaces \cite{a,b,c}. The common feature
of discrete and non-commutative models is the deformation of Leibnitz rules
for partial derivatives. 
In both cases  the transformation operators appear on the right-hand side
of formulae.
Their form is different, depending on the kind of model - in discrete models 
it is simply shift operator in given direction 
while on quantum spaces it is described by (\ref{mult},\ref{transf}).
Thus the main obstacle in extending the Takahashi-Umezawa method
is the fact that one of the 
operators on the right-hand side of (\ref{leibder}) acts simultaneously
on the first and second function 
in the product. In the discrete case we modified
Leibnitz rule using the inverse transformation 
operator which was simply the bac
k-shift operator
on the lattice. Investigating the special 
case of Klein-Gordon equation on quantum Minkowski 
spaces with $ Z=0 $ we also obtained the 
inverse operator $ \zeta^{-} $ , showed how it is connected with 
the transformation operator via 
$ * $-operation and applied it to modification of Leibnitz rule. \\
This modification is also possible in 
the general case which we now investigate. It is easy
to conjecture from the proof of 
Proposition 1.1 of \cite{o} that the operator:
\begin{equation}
\zeta^{- \;\; i}_{j}:= * \zeta^{i}_{j} *
\end{equation}
is the inverse transformation operator fulfilling:
\begin{equation}
\zeta^{- \;\; m}_{j} \zeta^{i}_{m}
= \zeta^{m}_{j} \zeta^{- \;\; i}_{m} =\delta^{i}_{j}
\label{inverse}
\end{equation}
The explicit form of the inverse 
operator for arbitrary function results from its
multiplicity property:
\begin{equation}
\zeta^{- \;\; i}_{j} (fg)
=(\zeta^{- \;\; m}_{j} f)(\zeta^{- \;\; i}_{m} g)
\end{equation}
and from its action on monomials of the first order:
\begin{equation}
\zeta^{- \;\; i}_{j} x^{k}
=R^{ki}_{\;\;\;\; ja} x^{a}+Z^{ki}_{\;\;\;\; j}
\end{equation}
Now using the formula 
(\ref{inverse}) and changing the product on the left-hand side
of Leibnitz rule (\ref{leibder}) we obtain its modification:
\begin{equation}
\partial^{i}[(\zeta^{- \;\; a}_{i} f)g]=(-\partial^{\dagger \;\; a} f)g
+f \partial^{a} g
\label{leibmod}
\end{equation}
where we use the following  notation for conjugated derivative
\begin{equation}
 \partial^{\dagger \;\; a}:
 =-\partial^{i} \zeta^{- \;\; a}_{i}=-\partial^{i} *\zeta^{a}_{i} * 
 \label{conjug}
\end{equation}
The conjugated derivative $ \partial^{\dagger} $ can be described
using the $ * $-operation. Let us notice that similarly 
to the special case from \cite{d}
the Leibnitz rules for 
conjugated derivative and the operator $ * (-\partial) * $
are identical:
\begin{eqnarray}
& \partial^{\dagger \;\; a} 
(fg)=(\partial^{\dagger \;\; i} f) \zeta^{- \;\; a}_{i}g
+f\partial^{\dagger \;\; a} g  &         \label{leibconj} \\
& *(-\partial^{a}) * (fg)= (-*\partial^{i} * f) \zeta^{- \;\; a}_{i}g +
f(-*\partial^{a} *g) &                   \label{leibstar}
\end{eqnarray}
It is easy to check that the 
action of both operators on monomials of the first order
is the same:
\begin{equation}
\partial^{\dagger \;\; a} x^{k}=-g^{ka}
\hspace{2cm} * (-\partial^{a} )* x^{k}=-g^{ka}
\end{equation}
To this aim we have used the 
definition of $ \partial^{\dagger } $ given in (\ref{conjug})
and the property of metric tensor 
$ \overline{g}^{ij}=g^{ji} $ from \cite{o}. \\
Now Leibnitz rules for both 
operators (\ref{leibconj}, 
\ref{leibstar}) and their equality for monomials of the first order
imply the identity of 
$ \partial^{\dagger \;\; a} $ and $ *(-\partial^{a}) * $
for arbitrary monomials
by virtue of mathematical 
induction principle. Therefore they 
act in the same way on all 
functions on quantum Minkowski space, thus are identical:
\begin{equation}
 \partial^{\dagger \;\; a} = *(-\partial^{a}) *
 \label{conjstar}
\end{equation}
\section{The conservation laws for linear equations of motion
on quantum Minkowski spaces}
\subsection{Equations of motion on quantum Minkowski space}
Lately a number of equations of motion on quantum spaces were studied
in the lierature. They 
include the Klein-Gordon  and Dirac equations and their solutions 
investigated by Podle\'{s} \cite{o} as well as equations considered 
on q-Minkowski space from \cite{g,h,i,j,k,t,u}
 built within the framework of braided
 differential calculus \cite{ab,m,n,l}.
 In addition some quantum models
 on non-commutative spaces, in which
 deformation of commutation relations
 is motivated by Heisenberg 
 principle and classical 
 gravity, were studied by Doplicher {\em et al.} in \cite{e,f}.
  \\
 In this section we shall 
 consider general, linear equation on quantum Minkowski space defined
 by (\ref{comx}). These operators 
 include the Klein-Gordon and Dirac operators from \cite{o}.
 We construct the operator of 
 equation using partial derivatives fulfilling (\ref{comp},\ref{com})
 in the following way:
 \begin{eqnarray}
  & \Lambda (\partial) \Phi=0 &     \label{equation} \\
  & \Lambda (\partial)=\Lambda_{0}
  + \sum_{l=1}^{N} \Lambda_{\mu_{1}... \mu_{l}} \label{lambda}
  \partial^{\mu_{1}}...\partial^{\mu_{l}} &
 \end{eqnarray}
 As the derivatives in (\ref{lambda}) 
 R-commute we have the following 
 property of the coefficients (which may be matrices)
 with respect to permutation of indices:
 \begin{equation}
 R^{\mu_{k} \mu_{k+1}}_{\;\;\;\;\;
 \;\;\;\;\;\;\;\;\; \nu \rho}
  \Lambda_{\mu_{1}...\mu_{k} \mu_{k+1}... \mu_{l}}=
 \Lambda_{\mu_{1}...\nu \rho... \mu_{l}}
 \end{equation}
 where $  l=1,...,N $ and
 and $k=1,...,l-1 $;
 what means that they are R-symmetric with respect to permutations. \\
 In analogy with  the classical 
 field theory we consider constant coefficients
 and this implies that they  obey the following equations:
 \begin{eqnarray}
 &   \zeta^{\mu_{k}}_{j} 
 \Lambda_{\mu_{1}...\mu_{k-1}\mu_{k}... \mu_{l}}
 =\Lambda_{\mu_{1}...\mu_{k-1}j... \mu_{l}} &  
    \label{const1} \\
&   \partial^{j} \Lambda_{\mu_{1}... \mu_{l}}=0 &      \label{const2} 
 \end{eqnarray}
 where $ l=1,...,N \hspace{0,5cm} k=1,...,l \hspace{0,5cm}
 j=1,2,3,4 $. \\
 Let us notice that formulae (\ref{inverse},\ref{const1}) imply also:
 \begin{equation}
 \zeta^{- \;\; \mu_{k}}_{j}
  \Lambda_{\mu_{1}...\mu_{k-1}\mu_{k}... \mu_{l}}
 =\Lambda_{\mu_{1}...\mu_{k-1}j... \mu_{l}} \
 \label{constinverse}
 \end{equation}
 for  $ l=1,...,N \hspace{0,5cm} k=1,...,l \hspace{0,5cm}
 j=1,2,3,4 $
 while (\ref{conjstar},
 \ref{const2}) and the 
 condition of $ * $-invariance of the equation operator
  (\ref{condstar}) yields for $ * $-invariant equations:
 \begin{equation}
\partial^{\dagger \;\; j} \Lambda_{\mu_{1}... \mu_{l}}=0 
\hspace{1cm} l=1,...,N \;\;\;\; j=1,2,3,4    
 \end{equation}
 \subsection{The operators $ \Gamma $ and $\hat{\Gamma}$ }
 In order to derive the 
 conservation law for equation (\ref{equation}) we need an operator
 $ \Gamma $ which in the classical procedure of Takahashi-Umezawa
 fulfills the equality:
 \begin{equation}
 \sum_{\mu}  ( \stackrel{\leftarrow}
 {\partial}^{\mu} +\partial^{\mu} )
 \Gamma_{\mu} (\partial, \stackrel{\leftarrow}{\partial})=
 \Lambda(\partial)- \Lambda(-\stackrel{\leftarrow}{\partial})
 \label{condg}
 \end{equation}
 In the above formula the 
 partial derivatives obey the rule of classical differential calculus,
so derivatives acting on the left- and right-hand side commute. \\
This is not the case in 
non-commutative  differential calculi, in which the derivatives
do not commute according 
to (\ref{comp}). Additionaly we shall deal with conjugated derivatives
introduced by modification 
of Leibnitz rule. The set of derivatives and their
conjugations in the sense of 
(\ref{conjug}) becomes 
commutative only in special case $ R=\tau $. \\
As we consider the general 
case we should replace the equality (\ref{condg}) with 
the following condition for the operator $ \Gamma $:
\begin{equation}
\sum_{\mu} (-\stackrel{\leftarrow}
{\partial}^{\dagger \;\; \mu} +\partial^{\mu} ) \circ
\Gamma_{\mu} (\partial, \stackrel{\leftarrow}{\partial}^{\dagger})=
 \Lambda(\partial)-
  \Lambda(\stackrel{\leftarrow}{\partial}^{\dagger})
 \label{cond}
\end{equation}
where the operator $
 \Lambda(\stackrel{\leftarrow}{\partial}^{\dagger}) $
looks as follows:
\begin{equation}
\Lambda(\stackrel{\leftarrow}{\partial}^{\dagger})= \Lambda_{0}
+ \sum_{l=1}^{N}
  \stackrel{\leftarrow}{\partial}^{\dagger \;\; \mu_{1}}...
  \stackrel{\leftarrow}{\partial}^{\dagger \;\; \mu_{l}}
   \Lambda_{\mu_{1}... \mu_{l}}
\label{lambdaconj}   
\end{equation}
We introduced the notation for the product "$\circ$" to underline
the way it acts on monomials of derivatives:
\begin{eqnarray}
 & (-\stackrel{\leftarrow}
 {\partial}^{\dagger \;\; \mu} +\partial^{\mu} ) \circ
\overline{[\nu_{1},...,\nu_{l}]}
a(\vec{x})[\rho_{1},...,\rho_{k}]:= & \nonumber \\
 & - \overline{[\nu_{1},...,\nu_{l},
 \mu]}a(\vec{x})[\rho_{1},...,\rho_{k}]
+\overline{[\nu_{1},...,\nu_{l}]}
\partial^{\mu} a(\vec{x})[\rho_{1},...,\rho_{k}] &
\end{eqnarray}
where we have denoted
the monomials of derivatives as follows:
\begin{eqnarray}
& [\rho_{1},...,\rho_{k}]:
=\partial^{\rho_{1}}... \partial^{\rho_{k}} &  \label{not1} \\
& \overline{[\nu_{1},...,\nu_{l}]}:=
\stackrel{\leftarrow}{\partial}^{\dagger \;\; \nu_{1}}...
\stackrel{\leftarrow}{\partial}^{\dagger \;\; \nu_{l}} & \label{not2}
\end{eqnarray}
\newtheorem{one}{Proposition}[section]
\begin{one}
The unique solution of 
(\ref{cond}) in class of polynomials of derivatives
$ \stackrel{\leftarrow}{\partial}^{\dagger} $ and $ \partial $
is of the form:
\begin{equation}
\Gamma_{\mu} (\partial, \stackrel{\leftarrow}{\partial}^{\dagger})=
\Lambda_{\mu} +\sum_{l=1}^{N-1} \sum_{k=0}^{l} 
\stackrel{\leftarrow}{\partial}^{\dagger \;\; \mu_{1}}...
\stackrel{\leftarrow}{\partial}^{\dagger \;\; \mu_{k}}
\Lambda_{\mu_{1}...\mu_{k} \mu \mu_{k+1}...\mu_{l}}
\partial^{\mu_{k+1}}... \partial^{\mu_{l}}
\end{equation}
\end{one}
Proof: \\
The technique we use is 
very similar to that applied in proof of analogous
proposition for discrete 
models \cite{c}. We denote the monomials of derivatives
as described above (\ref{not1},\ref{not2}). \\
Now the modified Leibnitz 
rule (\ref{leibmod}) implies that in order to solve (\ref{cond})
we should consider the general polynomial of order N-1 with functional 
coefficients of the following form:
\begin{equation}
\Gamma_{\mu} (\partial, \stackrel{\leftarrow}{\partial}^{\dagger})=
a^{0}_{\mu} +  \sum_{l=1}^{N-1} \sum_{k=0}^{l} 
\overline{[\mu_{1},...,\mu_{k}]}a^{k}_{\mu \mu_{1}... \mu_{l}}
[\mu_{k+1},...,\mu_{l}]
\label{solution}
\end{equation}
We apply the condition 
(\ref{cond}) to the general 
from of the solution (\ref{solution}) to derive
the equations for coefficients $ a^{k}_{\mu \mu_{1}... \mu_{l}}$:
\begin{equation}
\sum_{\mu} (-\stackrel{\leftarrow}
{\partial}^{\dagger \;\; \mu} +\partial^{\mu} ) \circ
\Gamma_{\mu} (\partial, 
\stackrel{\leftarrow}{\partial}^{\dagger})=
\end{equation}
$$ - \sum_{l=1}^{N-1} \sum_{k=0}^{l} \sum_{\mu}
\overline{[\mu_{1},...,\mu_{k},\mu]}a^{k}_{\mu \mu_{1}... \mu_{l}}
[\mu_{k+1},...,\mu_{l}]  $$
$$ +  \sum_{l=1}^{N-1} \sum_{k=0}^{l} 
\overline{[\mu_{1},...,\mu_{k}]}
\sum_{\mu}(\zeta^{\mu}_{\nu}a^{k}_{\mu \mu_{1}... \mu_{l}})
[\nu,\mu_{k+1},...,\mu_{l}] $$
$$  + \sum_{l=1}^{N-1} \sum_{k=0}^{l} 
\overline{[\mu_{1},...,\mu_{k}]}
\sum_{\mu} (\partial^{\mu}a^{k}_{\mu \mu_{1}... \mu_{l}})
[\mu_{k+1},...,\mu_{l}] $$
$$ -\sum_{\mu}\overline{[\mu]}a^{0}_{\mu}
+ \sum_{\mu} (\zeta^{\mu}_{\nu} a^{0}_{\mu} )[\nu]
+\sum_{\mu} (\partial^{\mu} a^{0}_{\mu}) = $$
$$ =\Lambda(\partial)- 
\Lambda(\stackrel{\leftarrow}{\partial}^{\dagger}) $$
We compare the coefficients 
of monomials of the same type on both sides of the above condition
and obtain the following set 
of equations for functions $ a^{k}_{\mu \mu_{1}... \mu_{l}} $:
\begin{eqnarray}
& a^{0}_{\mu}= \Lambda_{\mu}  & \label{eq0} \\
& \partial^{\alpha}a^{0}_{\alpha \mu}
+\zeta^{\nu}_{\mu} a^{0}_{\nu}=\Lambda_{\mu}  & \label{eq1}  \\
& \zeta^{\alpha}_{\mu}a^{0}_{\alpha \mu_{1}... \mu_{l}}
+ \partial^{\alpha} a^{0}_{\alpha \mu \mu_{1} ...\mu_{l}}= 
\Lambda_{ \mu \mu_{1}...\mu_{l}} &    \label{eq2} \\
 & -a^{k}_{\mu \mu_{1}...\mu_{l}}
+ \zeta^{\alpha}_{\mu_{k+1}} 
a^{k+1}_{\alpha \mu_{1} ... \mu_{k} \mu \mu_{k+2}... \mu_{l}}
+ \partial^{\alpha} a^{k+1}_{\alpha \mu_{1}... 
\mu_{k} \mu \mu_{k+1}... \mu_{l}}=
0 &   \label{eq3} \\
& a^{l}_{\mu \mu_{1}... \mu_{l}}+
 \partial^{\alpha} a^{l}_{\alpha \mu_{1}...\mu_{l}\mu}=
\Lambda_{\mu_{1}...\mu_{l} \mu} & \label{eq4}
\end{eqnarray}
with $ l=1,...,N-1 \hspace{1cm}  k=0,...,l-1 $. \\
Starting from (\ref{eq0},\ref{eq1},{\ref{eq2}) 
we get the unique solution for coefficients $ a^{0} $
\begin{equation}
a^{0}_{\mu}=\Lambda_{\mu} 
\hspace{2cm} a^{0}_{\mu \mu_{1}...
 \mu_{l}}=\Lambda_{\mu \mu_{1}... \mu_{l}} \hspace{1cm} l=1,...,N-1
\label{solzero}
\end{equation}
This R-symmetric, constant in 
the sense of (\ref{const1},\ref{const2}) 
solution for initial coefficients
allows us to evaluate the remaining ones using (\ref{eq3}). \\
We start from  equations for $ l=N-1$ and conclude that they reduce to
the following set:
\begin{equation}
-a^{k}_{\mu \mu_{1}...\mu_{N-1}}
+ \zeta^{\alpha}_{\mu_{k+1}} 
a^{k+1}_{\alpha \mu_{1} ... \mu_{k} \mu \mu_{k+2}... \mu_{N-1}}=0
\hspace{1cm} 0\leq  k<N-1
\label{set}
\end{equation}
As we know the explicit expressions 
for $ a^{0}_{ \mu_{1}...\mu_{N}}  $ from (\ref{solzero})
we are able to derive from the above equation 
and from the properties of coefficients of 
the equation (\ref{const1},\ref{constinverse})
  the solution for
$  a^{1}_{\mu_{1} ... \mu_{N}} $ which 
has the following form:
\begin{equation}
 a^{1}_{\mu \mu_{1} \mu_{2} ... 
 \mu_{N-1}}=\Lambda_{\mu_{1} \mu \mu_{2}...\mu_{N-1}}
 \label{sol1}
 \end{equation}
Using the explicit form of coefficients 
of type $ a^{1}_{\mu_{1}...\mu_{N}} $ 
we perform the next step of solution 
of the set (\ref{set}) and solve it for $ k=1 $
applying the same method as before, namely rewriting the set
of equations, putting the solution (\ref{sol1})
into it and then using (\ref{constinverse}). 
After subsequent calculations
for $ k=1,...,N-1 $ we obtain the general form
of coefficients $ a^{k}_{\mu_{1}...\mu_{N}} $ :
\begin{equation}
a^{k}_{\mu \mu_{1}...\mu_{N-1}}=
\Lambda_{\mu_{1}...\mu_{k} \mu \mu_{k+1}...\mu_{N-1}}
\label{solN}
\end{equation}
Now from the form of coefficients 
$ a^{k}_{\mu_{1}...\mu_{N}} $ we conclude
that they all fulfill the condition 
(\ref{const2}) so for $ l=N-2 $
we also obtain the set of equations 
in which the part with the full
divergence vanishes:
\begin{equation}
-a^{k}_{\mu \mu_{1}...\mu_{N-2}}
+ \zeta^{\alpha}_{\mu_{k+1}} 
a^{k+1}_{\alpha \mu_{1} ...
 \mu_{k} \mu \mu_{k+2}... \mu_{N-2}}=0
\hspace{1cm} 0\leq  k<N-2
\label{set1}
\end{equation}
The method of solving equations 
(\ref{set1}) follows 
the calculations done for $ l=N-1 $. We start
from the known coefficient of type $ a^{0} $
described by (\ref{solzero}) and derive the remaining ones
using the subsequent equations from (\ref{set1}). \\
The result as before are expressions 
constant in the sense of (\ref{const1},\ref{const2})
of the  form:
\begin{equation}
a^{k}_{\mu \mu_{1}...\mu_{N-2}}=
\Lambda_{\mu_{1}...\mu_{k}\mu \mu_{k+1}... \mu_{N-2}}
\end{equation}
It is obvious that the next steps 
are analogous so we shall not present them
in detail. \\
The general and unique solution of 
the set (\ref{eq0} - \ref{eq4})
looks as follows:
\begin{equation}
a^{k}_{\mu \mu_{1}...\mu_{l}}
=\Lambda_{\mu_{1}...\mu_{k} \mu \mu_{k+1}... \mu_{l}}
\hspace{1cm} l=1,...,N-1 \;\;\;\; k=0,...,l
\end{equation}
The derivation of the explicit 
formulae for unique solution of the coefficients 
of the operator $ \Gamma_{\mu} $
 concludes the proof of Proposition 3.1. \\
\\
From the above proof of 
Proposition 3.1 we conclude that the unique solution
of the equation (\ref{cond}) can  also be derived for equations
in which the coefficients obey the requirement (\ref{const1}),
but the condition (\ref{const2}) is  weakend as follows:
\begin{equation}
 \sum_{\mu_{k}} \partial^{\mu_{k}} 
 \Lambda_{\mu_{1}...\mu_{k}...\mu_{l}} =0
\hspace{2cm} l=1,..,N \;\;\;\; k=1,...,l
\label{constweak}
\end{equation}
\newtheorem{cor1}[one]{Corollary}
\begin{cor1}
The unique solution of (\ref{cond}) 
in class of polynomials of derivatives
$ \stackrel{\leftarrow}{\partial}^{\dagger} $ and $ \partial $
for the equation operator $
 \Lambda $ fulfilling (\ref{const1},\ref{constweak})
is of the form:
\begin{equation}
\Gamma_{\mu} (\partial, \stackrel{\leftarrow}{\partial}^{\dagger})=
\Lambda_{\mu} +\sum_{l=1}^{N-1} \sum_{k=0}^{l} 
\stackrel{\leftarrow}{\partial}^{\dagger \;\; \mu_{1}}...
\stackrel{\leftarrow}{\partial}^{\dagger \;\; \mu_{k}}
\Lambda_{\mu_{1}...\mu_{k} \mu \mu_{k+1}...\mu_{l}}
\partial^{\mu_{k+1}}... \partial^{\mu_{l}}
\end{equation}
\end{cor1}
In contrast with classical case where it is sufficient to know
the operator $ \Gamma_{\mu} $ to construct the conserved currents, 
we must additionally modify $ \Gamma_{\mu} $ due to the
deformation of Leibnitz rule (\ref{leibder}). \\
We introduce the operator $ \hat{\Gamma}_{\mu} $ in the form:
\begin{equation}
\hat{\Gamma}_{\mu} 
(\partial, \stackrel{\leftarrow}{\partial}^{\dagger})=
\stackrel{\leftarrow}{\zeta}^{- \;\; j}_{\mu}\Lambda_{j}
 +\sum_{l=1}^{N-1} \sum_{k=0}^{l} 
\stackrel{\leftarrow}{\partial}^{\dagger \;\; \mu_{1}}...
\stackrel{\leftarrow}{\partial}^{\dagger \;\; \mu_{k}}
\stackrel{\leftarrow}{\zeta}^{- \;\; j}_{\mu}
\Lambda_{\mu_{1}...\mu_{k} j \mu_{k+1}...\mu_{l}}
\partial^{\mu_{k+1}}... \partial^{\mu_{l}}
\label{gammahat}
\end{equation}
As we see the modification consists of introducing the 
inverse transformation operator
 $ \zeta^{-} $ in the monomials between derivatives 
acting on the left- and right-hand side.
\subsection{The conservation laws and conserved currents}
In this section we derive the conservation laws for
linear  equations with 
constant coefficients  
described by (\ref{equation},\ref{lambda}). To this aim
it is necessary to have 
the solutions of 
initial equation and of its conjugation (\ref{conjug}).
Having these two functions we can formulate
the following 
proposition which describes the explicit form of conserved 
currents for equation (\ref{equation}).
\newtheorem{two}[one]{Proposition}
\begin{two}
Let us assume that function 
$ \Phi $ is an arbitrary solution of equation (\ref{equation})
with coefficients fulfilling (\ref{const1},\ref{const2}), that means:
\begin{equation}
\Lambda (\partial) \Phi =0
\end{equation}
and function $ F $ solves the conjugated equation:
\begin{equation}
F \Lambda (\stackrel{\leftarrow}{\partial}^{\dagger}) =0
\end{equation}
Then 
\begin{equation}
J_{\mu}=F \hat{\Gamma}_{\mu}
 (\partial, \stackrel{\leftarrow}{\partial}^{\dagger})\Phi 
\end{equation}
where the operator
 $ \hat{\Gamma}_{\mu} $ 
  is defined by (\ref{gammahat}), is a current which
obeys the conservation law
 in given differential calculus on quantum Minkowski
space:
\begin{equation}
\sum_{\mu} \partial^{\mu} J_{\mu}=0
\end{equation}
\end{two}
Proof: \\
In the straightforward proof of the proposition we use the modified
Leibnitz rule (\ref{leibmod})
 and the properties of operator $ \Gamma_{\mu} $ (\ref{cond})
as well as properties of 
coefficients of equation (\ref{const1},\ref{const2}):
\begin{equation}
\sum_{\mu} \partial^{\mu} J_{\mu}=
\end{equation}
$$
\sum_{\mu} \partial^{\mu} F 
\hat{\Gamma}_{\mu} (\partial, 
\stackrel{\leftarrow}{\partial}^{\dagger})\Phi =
 \sum_{\mu} \partial^{\mu} F  (
\stackrel{\leftarrow}{\zeta}^{- \;\; j}_{\mu}\Lambda_{j}+ $$
$$
 +\sum_{l=1}^{N-1} \sum_{k=0}^{l} 
\stackrel{\leftarrow}{\partial}^{\dagger \;\; \mu_{1}}...
\stackrel{\leftarrow}{\partial}^{\dagger \;\; \mu_{k}}
\stackrel{\leftarrow}{\zeta}^{- \;\; j}_{\mu}
\Lambda_{\mu_{1}...\mu_{k} j \mu_{k+1}...\mu_{l}}
\partial^{\mu_{k+1}}... \partial^{\mu_{l}}  ) \Phi = $$
$$ \sum_{j} F \left (
 (-\stackrel{\leftarrow}{\partial}^{\dagger \;\; j}
\Lambda_{j} +\Lambda_{j} \partial^{j}) \right ) \Phi + $$
 $$ + \sum_{j} F   (\sum_{l=1}^{N-1} \sum_{k=0}^{l} 
\stackrel{\leftarrow}{\partial}^{\dagger \;\; \mu_{1}}...
\stackrel{\leftarrow}{\partial}^{\dagger \;\; \mu_{k}}
(-\stackrel{\leftarrow}{\partial}^{\dagger \;\; j}
\Lambda_{\mu_{1}...\mu_{k} j \mu_{k+1}...\mu_{l}} +
$$
$$
+ \Lambda_{\mu_{1}...
\mu_{k} j \mu_{k+1}...\mu_{l}} \partial^{j})
\partial^{\mu_{k+1}}... \partial^{\mu_{l}}  ) \Phi =  $$
$$ F \left (\sum_{j}
 (-\stackrel{\leftarrow}
 {\partial}^{\dagger \;\; j} + \partial^{j})
\circ 
\Gamma_{\mu} (\partial, 
\stackrel{\leftarrow}{\partial}^{\dagger}) \right ) \Phi= $$
$$ F \left ( \Lambda 
(\partial)-\Lambda( 
\stackrel{\leftarrow}{\partial}^{\dagger})
\right ) \Phi=0 $$
Thus the conservation law for arbitrary linear equation with constant 
coefficients is valid provided functions 
$ F $ and $ \Phi $ are solutions of corresponding equations. \\
\\
\newtheorem{cor2}[one]{Corollary}
\begin{cor2}
If function $ \Phi $ is 
an arbitrary solution of equation (\ref{equation})
with coefficients fulfilling (\ref{const1},\ref{constweak}):
$$ \Lambda(\partial) \Phi=0 $$
and function $ F $ solves its conjugation (\ref{conjug}):
$$ F \Lambda( \stackrel{\leftarrow}{\partial}^{\dagger} )=0 $$
then the current of the form:
$$ J_{\mu}=F \hat{\Gamma}_{\mu}
 (\partial, \stackrel{\leftarrow}{\partial}^{\dagger})\Phi  $$
with $ \hat{\Gamma} $ given by (\ref{gammahat}) is conserved:
$$ \sum_{\mu} \partial^{\mu} J_{\mu}=0 $$
\end{cor2}
Proof: \\
It results from the proof 
of the Proposition 3.3 and from Corollary 3.2
describing operator fulfilling (\ref{cond}) for equations 
obeying weakend condition (\ref{const1},\ref{constweak}). \\
\\
Let us observe that in the special case of Klein-Gordon equation on
quantum  Minkowski space with  
$ Z=0 $ we were able to connect the solution of conjugated
equation $ F $ with the  
$ * $-transformation of 
$ \Phi $ \cite{d}. This possibility was due to
the fact that Klein-Gordon 
operator is an $*$-invariant one \cite{o}.
\\
We shall now check the conditions 
of   $ * $-invariance for the 
operator of equation of the form (\ref{equation},\ref{lambda}).
Taking into account $ (i\partial)^{*}=i\partial $ from \cite{o}
we see that after $ * $-operation we obtain:
\begin{equation}
\Lambda(\partial)^{*} =
\Lambda_{0}^{*}+
\sum_{l=1}^{N} (\partial^{\mu_{l}})^{*}...
(\partial^{\mu_{1}})^{*}
\Lambda_{\mu_{1}...\mu_{l}}^{*}=
\end{equation}
$$ =\Lambda_{0}^{*}
+\sum_{l=1}^{N} (-1)^{l}\partial^{\mu_{l}}...
\partial^{\mu_{1}}
\Lambda_{\mu_{1}...\mu_{l}}^{*}=\Lambda(\partial) $$
Comparing the coefficients of the $ \Lambda(\partial)^{*} $ with
coefficients of the initial operator $ \Lambda(\partial) $
we conclude that the following proposition is valid
\newtheorem{three}[one]{Proposition }
\begin{three}
The operator of equation 
(\ref{equation},\ref{lambda}) is $ * $-invariant:
\begin{equation}
\Lambda(\partial)^{*}=\Lambda(\partial)
\end{equation}
iff the coefficients fulfill the conditions:
\begin{equation}
\Lambda^{*}_{0}=\Lambda_{0} \hspace{2cm}
\Lambda_{\mu_{1}...\mu_{l}}^{*}
=\Lambda_{\mu_{l}...\mu_{1}} (-1)^{l}
\hspace{1cm} l=1,...,N
\label{condstar}
\end{equation}
\end{three}
For equations fulfilling 
$ * $-invariance condition we can express the solution
of conjugated 
equation (\ref{lambdaconj}) in terms of solutions of initial 
equation (\ref{equation},\ref{lambda}).
Therefore the conserved 
currents for such equations can be constructed using only the 
latter solution.
\newtheorem{four}[one]{Proposition}
\begin{four}
For linear equation 
with constant coefficients fulfilling the conditions 
(\ref{const1},\ref{const2},\ref{condstar})
the current of the form :
\begin{equation}
J_{\mu}=\Phi^{*} 
\hat{\Gamma}_{\mu} (\partial, 
\stackrel{\leftarrow}{\partial}^{\dagger})\Phi 
\label{currenth}
\end{equation}
where $ \Phi $ is an arbitrary 
solution of (\ref{equation},\ref{lambda}) is conserved:
\begin{equation}
\sum_{\mu} \partial^{\mu} J_{\mu}=0
\end{equation}
\end{four}
Proof: \\
The Proposition 3.3 implies that we need only to show:
\begin{equation}
\Phi^{*} \Lambda 
(\stackrel{\leftarrow}{\partial}^{\dagger})=0
\end{equation}
provided $ \Phi $ is an 
arbitrary solution of $ \Lambda(\partial) \Phi=0 $. 
Let us check it using 
the property of conjugated derivative (\ref{conjstar}):
\begin{equation}
\Phi^{*} \Lambda 
(\stackrel{\leftarrow}{\partial}^{\dagger})=
\end{equation} 
$$ \sum_{l=0}^{N} 
\Phi^{*}\stackrel{\leftarrow}
{\partial}^{\dagger \;\; \mu_{l}}...
\stackrel{\leftarrow}{\partial}^{\dagger \;\; \mu_{1}}
\Lambda_{\mu_{1}...\mu_{l}}= $$
$$ \sum_{l=0}^{N} \Phi^{*}
 (-1)^{l} \stackrel{\leftarrow}{*} 
\stackrel{\leftarrow}{\partial}^{ \mu_{l}}...
\stackrel{\leftarrow}
{\partial}^{ \mu_{1}} \stackrel{\leftarrow}{*}
\Lambda_{\mu_{1}...\mu_{l}}= $$

$$ \sum_{l=0}^{N} \stackrel{\leftarrow}{*}
(-1)^{l} \Lambda_{\mu_{l}...\mu_{1}} \Phi (-1)^{l} 
\stackrel{\leftarrow}{\partial}^{ \mu_{l}}...
\stackrel{\leftarrow}{\partial}^{ \mu_{1}} 
 \stackrel{\leftarrow}{*}=0 $$
\\
When two solutions 
of equation of motion (\ref{equation}) are known 
one can construct the current according to the following corollary
which is the result of Propositions 3.3 and 3.6:
\newtheorem{cor3}[one]{Corollary}
\begin{cor3}
For linear equation 
with coefficients fulfilling 
(\ref{const1},\ref{const2},\ref{condstar})
the current of the form:
\begin{equation}
J_{\mu}=i (\Phi ')^{*} 
\hat{\Gamma}_{\mu} 
(\partial, \stackrel{\leftarrow}{\partial}^{\dagger})\Phi 
-i \Phi^{*} 
\hat{\Gamma}_{\mu} 
(\partial, \stackrel{\leftarrow}{\partial}^{\dagger})\Phi ' 
\label{currentc}
\end{equation}
where $ \Phi $ and $ 
\Phi ' $ are arbitrary solutions of (\ref{equation})
is conserved:
\begin{equation}
 \sum_{\mu} \partial^{\mu} J_{\mu}=0   \label{star}
\end{equation}
\end{cor3}
\section{Applications}
We have developed simple method of derivation of conserved currents
for linear equations on 
a class of quantum Minkowski spaces. This procedure can 
be applied to equations with 
coefficients constant in the sense of (\ref{const1},\ref{const2})
or fulfilling weakend conditions (\ref{const1},\ref{constweak}).
Now we shall apply the presented technique
to a few equations on different quantum Minkowski spaces. 
Some of these equations were studied earlier in \cite{o} where
also their solutions were constructed. \\
Following the 
classical field theory we shall obtain different solutions
of equation of motion using the symmetry transformation operators. 
In the examples 
we show that they are quantum deformations of classical
 operators plus 
 the transformation operator (\ref{mult},\ref{transf}).
 The algebraic and 
 possible co-algebraic 
 properties of the set of symmetry operators
 are still to be 
 investigated, nevertheless we wish to point out
 that in the special case studied 
 in \cite{d} they form closed algebra and we hope
 to obtain their co-algebraic 
 structure from Leibnitz rules of symmetry operators
 determined by Leibnitz rules for 
 derivatives and variables (\ref{leibder},\ref{leibx}).
\subsection{Klein-Gordon equation}
Klein-Gordon equation on quantum 
Minkowski space in the sense of 
(\ref{comx}) was introduced by Podle\'{s} in \cite{o}
where also its solutions were studied. It looks as follows:
\begin{equation}
(\Box+m^{2})\Phi =0
\label{kg}
\end{equation}
with d'Alembert's operator built 
using exterior partial derivatives from non-com- \\
mutative
differential calculus
(\ref{comp},\ref{com},\ref{comx}):
\begin{equation}
\Box=g_{ab}\partial^{a} 
\partial^{b}=\partial^{a} \partial^{b}g_{ab}
\end{equation}
The consistency conditions which allow us to write coefficients
of equation in front or after the differential operators coincide
with requirements studied 
in the previous section  (\ref{const1},\ref{const2}). \\
In our construction we shall 
consider currents connected with symmetry transform
ations
of solutions of Klein-Gordon equation. The special case
for $ Z=0 $ was solved earlier 
in \cite{d} where also algebraic properties
of symmetry transformation 
operators were investigated. Let us now assume
that $ R $, $ Z $ and $ T $ are 
arbitrary matrices and tensors allowed in calculus
on quantum Minkowski space and check the commutator of d'Alembert's
operator with variable $ x^{k} $ :
\begin{equation}
[\Box, x^{k}]=2 \partial^{k}
\end{equation}
Taking into account the property (\ref{comp}) rewritten as follows:
$$ (R-1)^{kl}_{\;\;\;\; ab} \partial^{a} \partial^{b}=0 $$
we can easily construct the symmetry transformation operator
analogous to angular momentum operator from classical field theory:
\begin{equation}
M^{kl}=i (R-1)^{kl}_{\;\;\;\; ab} x^{a} \partial^{b}
\label{momentum}
\end{equation}
As these operators commute with 
Klein-Gordon equation operator they transform solution
of Klein-Gordon equation into another solution. \\
The set of symmetry operators can be 
also completed with momentum operators \cite{o}:
\begin{equation}
P^{l}=i\partial^{l}
\end{equation}
In addition, using properties 
of the transformation operator $ \zeta $
given in Proposition 3.1 of \cite{o}
one concludes that there is 
an additional symmetry operator for Klein-Gordon
equation:
\begin{equation}
[\Box, \zeta^{a}_{b}] =0
\end{equation}
due to the property of $ R $-matrix:
$$ g^{jb} R^{dc}_{\;\;\;\;ba}=R^{jd}_{\;\;\;\; ak}g^{kc} $$
Now we should construct the $ 
\hat{\Gamma} $ operator for our equation
using the general formula 
(\ref{gammahat}). It has the form identical with the one obtained
in \cite{d}:
\begin{equation}
\hat{\Gamma}_{\mu} 
(\partial,\stackrel{\leftarrow}{\partial}^{\dagger})=
\stackrel{\leftarrow}{\partial}^{\dagger \;\;a}g_{aj} 
\stackrel{\leftarrow}{\zeta}^{- \;\; j}_{\mu} 
+\stackrel{\leftarrow}{\zeta}^{- \;\; j}_{\mu} g_{ja} \partial^{a}
\label{gammakg}
\end{equation} 
The results (\ref{currentc},\ref{star}) 
presented earlier imply that the currents:
\begin{eqnarray}
& J^{kl}_{\mu}=
i \Phi^{*}\hat{\Gamma}_{\mu} 
(\partial,\stackrel{\leftarrow}{\partial}^{\dagger}) M^{kl} \Phi
-i (M^{kl} \Phi)^{*} \hat{\Gamma}_{\mu} 
(\partial,\stackrel{\leftarrow}{\partial}^{\dagger}) \Phi & \\
& J^{l}_{\mu}=
i \Phi^{*}\hat{\Gamma}_{\mu} 
(\partial,\stackrel{\leftarrow}{\partial}^{\dagger}) P^{l} \Phi
-i (P^{l} \Phi)^{*}
 \hat{\Gamma}_{\mu} (\partial,
 \stackrel{\leftarrow}{\partial}^{\dagger}) \Phi & \\
& J^{a}_{\mu \;\; b}=
i \Phi^{*}\hat{\Gamma}_{\mu} 
(\partial,\stackrel{\leftarrow}
{\partial}^{\dagger}) \zeta^{a}_{b} \Phi
-i (\zeta^{a}_{b} \Phi)^{*} 
\hat{\Gamma}_{\mu} (\partial,
\stackrel{\leftarrow}{\partial}^{\dagger}) \Phi &
\end{eqnarray}
are conserved  quantities by virtue of Corollary 3.7:
\begin{equation}
\sum_{\mu} \partial^{\mu} J^{kl}_{\mu}=0
\hspace{2cm} \sum_{\mu} J^{k}_{\mu}=0
\hspace{2cm} \sum_{\mu} \partial^{\mu} J^{a}_{\mu \;\; b}=0
\end{equation}
\subsection{Dirac equation on quantum Minkowski space with $ R= \tau $}
Let us remind the Dirac equation from \cite{o}:
\begin{equation}
(i\gamma_{\mu} \partial^{\mu} +m) \Psi=0
\label{dirac}
\end{equation}
In our case when $ 
R=\tau $ the $ \gamma $ 
matrices fulfill the following condition:
\begin{equation}
\gamma^{\mu} 
\gamma^{\nu}+\gamma^{\nu} \gamma^{\mu}=2g^{\mu \nu}
\end{equation}
Similarly to the Klein-Gordon operator the coefficients $ \gamma_{\mu} $
(which are now matrices) 
obey the consistency conditions (\ref{const1},\ref{const2}). \\
For the Dirac equation we know symmetry operators analogous   
to momentum from \cite{o}:
\begin{equation}
P^{l}=i \partial^{l}
\label{symd1}
\end{equation}
In order to construct the 
angular momentum operator we check the commutator of the Dirac
operator with the scalar angular momentum (\ref{momentum}):
\begin{equation}
[i\gamma_{\mu} \partial^{\mu}, M^{kl}]=
i\gamma_{\rho}g^{\rho l}
\partial^{k}-i\gamma_{\rho}g^{\rho k}\partial^{l}
+i\gamma_{\rho} Z^{l \rho}_{\nu} \partial^{\nu} \partial^{k}
-i\gamma_{\rho} Z^{k \rho}_{\nu} \partial^{\nu} \partial^{l}
\end{equation}
The additional term 
appearing on the 
right-hand side implies that the spinorial part of the
angular momentum must be extended:
\begin{equation}
M^{kl}_{\rm{spin}}=ix^{k}\partial^{l} -ix^{l} \partial^{k}
+\frac{1}{2}[\gamma^{k},\gamma^{l}]
+\frac{1}{2}(Z^{kj}_{\rho}
 \partial^{l} 
 -Z^{lj}_{\rho} \partial^{k})\gamma_{j}\gamma^{\rho}
\label{symd2}
\end{equation}
This symmetry 
operator contains the scalar part and spin part built using 
Dirac matrices 
of identical form as in the classical commutative models and
is extended 
with part depending on $ Z $ matrices due to commutation 
rule (\ref{com}). \\
Following the 
general method we construct the $ \hat{\Gamma} $ operator 
for (\ref{dirac}) using (\ref{gammahat}) :
\begin{equation}
\hat{\Gamma}_{\mu}
 (\partial,\stackrel{\leftarrow}{\partial}^{\dagger})=
i\stackrel{\leftarrow}{\zeta}^{- \;\; j}_{\mu} \gamma_{j}
\end{equation}
Thus the currents 
built using symmetry operators (\ref{symd1},\ref{symd2})
according to the Corollary 3.7:
\begin{eqnarray}
& J^{kl}_{\mu}=i 
 \Psi^{*} \stackrel{\leftarrow}{\zeta}^{- \;\; j}_{\mu}
  \gamma_{j} M^{kl} \Psi
-i (M^{kl} \Psi)^{*} 
\stackrel{\leftarrow}{\zeta}^{- \;\; j}_{\mu}
 \gamma_{j} \Psi & \\
& J^{l}_{\mu}=i \Psi^{*}
 \stackrel{\leftarrow}
 {\zeta}^{- \;\; j}_{\mu} \gamma_{j}  P^{l} \Psi
-i (P^{l} \Psi)^{*} 
\stackrel{\leftarrow}{\zeta}^{- \;\; j}_{\mu} \gamma_{j} \Psi &
\end{eqnarray}
are conserved:
\begin{equation}
\sum_{\mu} \partial^{\mu} J^{kl}_{\mu}=0
\hspace{2cm} \sum_{\mu} J^{k}_{\mu}=0
\end{equation}
Let us notice that we have not included the symmetry operator
connected wih the 
transformation operator because in the case when $ R=\tau $
it can be expressed by momenta (\ref{symd1}) \cite{y}.
\subsection{Wave equation on quantum Minkowski spaces with $ Z=0 $}
Let us now assume that the mass in the Klein-Gordon equation 
(\ref{kg}) is equal to zero. The result is the wave equation of the form:
\begin{equation}
\Box \Phi=0
\label{wave}
\end{equation}
We shall study the 
symmetry operators for class of Minkowski spaces (\ref{comx})
with $ Z=0 $. 
Similarly to the classical 
field theory the set of symmetry operators of the Klein-Gordon
equation can be extended by additional operators \cite{aa}. \\
It is  easy to check the following commutation relations:
\begin{equation}
[\Box,\frac{1}{2} g_{ab}(R^{ab}_{\;\;\;\; kl}+
\delta^{ab}_{\;\;\;\; kl})x^{k}\partial^{l}]=2 \Box
\end{equation}
\begin{equation}
[\Box,\frac{1}{2} g_{ab}(R^{ab}_{\;\;\;\; kl}+
\delta^{ab}_{\;\;\;\; kl})x^{k}x^{l}]
=2g_{ab}g^{ab}+2 g_{ab}(R^{ab}_{\;\;\;\; kl}
+\delta^{ab}_{\;\;\;\; kl})x^{k}\partial^{l}
\end{equation}
Let us denote:
\begin{equation}
D:=\frac{i}{2}  g_{ab}(R^{ab}_{\;\;\;\; kl}+
\delta^{ab}_{\;\;\;\; kl})x^{k}\partial^{l}
\label{dilatation}
\end{equation}
\begin{equation}
\hat{x}^{2}:=\frac{1}{2} g_{ab}(R^{ab}_{\;\;\;\; kl}+
\delta^{ab}_{\;\;\;\; kl})x^{k}x^{l}
\end{equation}
These operators allow us to 
construct additional symmetry operators, namely: \\
- the dilatation operator $ D $ given by (\ref{dilatation}) \\
- the conformal boosts $ K^{m} $:
\begin{equation}
K^{m}=i\hat{x}^{2} \partial^{m} -2Dx^{m}
\end{equation}
Acting on arbitrary solution of the wave equation (\ref{wave})
they produce another solution of this equation:
$$ \Box D \Phi=(D+2i)\Box \Phi=0 $$
$$ \Box K^{m} \Phi =(K^{m}-2ix^{m}) \Box \Phi=0 $$
The $ \hat{\Gamma}_{\mu}$ 
operator for wave equation is given by (\ref{gammakg})
$$ \hat{\Gamma}_{\mu} 
(\partial,\stackrel{\leftarrow}{\partial}^{\dagger})=
\stackrel{\leftarrow}{\partial}^{\dagger \;\;a}g_{aj} 
\stackrel{\leftarrow}{\zeta}^{- \;\; j}_{\mu} 
+\stackrel{\leftarrow}{\zeta}^{- \;\; j}_{\mu} g_{ja} \partial^{a}
$$
Using this operator and symmetry operators
from the set:
\begin{eqnarray}
 & P^{l}=i\partial^{l} & \\
 & M^{kl}=i(R-1)^{kl}_{\;\;\;\; ab}x^{a}\partial^{b} & \\
 & \zeta^{a}_{b} & \\
 & D=\frac{i}{2} 
 g_{ ab}(R+1)^{ab}_{\;\;\;\; kl}x^{k} \partial^{l} & \\
 & K^{m}=i\hat{x}^{2} \partial^{m}-2Dx^{m} & 
\end{eqnarray}
we can construct the 
full set of conserved currents for wave equation:
\begin{eqnarray}
& J^{kl}_{\mu}=
i \Phi^{*}\hat{\Gamma}_{\mu} 
(\partial,\stackrel{\leftarrow}{\partial}^{\dagger}) M^{kl} \Phi
-i (M^{kl} \Phi)^{*} 
\hat{\Gamma}_{\mu} 
(\partial,\stackrel{\leftarrow}{\partial}^{\dagger}) \Phi & \\
& J^{l}_{\mu}=
i \Phi^{*}\hat{\Gamma}_{\mu}
 (\partial,\stackrel{\leftarrow}{\partial}^{\dagger}) P^{l} \Phi
-i (P^{l} \Phi)^{*}
 \hat{\Gamma}_{\mu} (\partial,
 \stackrel{\leftarrow}{\partial}^{\dagger}) \Phi & \\
& J^{a}_{\mu \;\; b}=
i \Phi^{*}\hat{\Gamma}_{\mu}
 (\partial,\stackrel{\leftarrow}
 {\partial}^{\dagger}) \zeta^{a}_{b} \Phi
-i (\zeta^{a}_{b} \Phi)^{*}
 \hat{\Gamma}_{\mu} (\partial,
 \stackrel{\leftarrow}{\partial}^{\dagger}) \Phi & \\
& J^{D}_{\mu}=
i \Phi^{*}\hat{\Gamma}_{\mu}
 (\partial,\stackrel{\leftarrow}{\partial}^{\dagger}) D \Phi
-i (D \Phi)^{*} \hat{\Gamma}_{\mu}
 (\partial,\stackrel{\leftarrow}{\partial}^{\dagger}) \Phi & \\
& \tilde{J}^{l}_{\mu}=
i \Phi^{*}\hat{\Gamma}_{\mu}
 (\partial,\stackrel{\leftarrow}
 {\partial}^{\dagger}) K^{l} \Phi
-i (K^{l} \Phi)^{*} \hat{\Gamma}_{\mu} 
(\partial,\stackrel{\leftarrow}{\partial}^{\dagger}) \Phi & 
\end{eqnarray} 
which are conserved  according to Corollary 3.7. \\
Similarly to Klein-Gordon and Dirac eqauation in the case $ R=\tau $ the 
transformation operator $ \zeta $ is an symmmetry operator
described by momenta, so it can be then excluded from the set of independent 
symmetry operators.

\section{Final remarks}
The presented extension 
of Takahashi-Umezawa procedure gives explicit formulae
for construction of 
conserved currents for linear equations of motion on quantum
Minkowski spaces. There 
is an interesting technical analogy between non-commutative
differential calculus and 
discrete calculus on commutative spaces - the appearance
of the transformation operator 
in Leibnitz rules, which was the main obstacle in our construction. \\
For all presented equations 
this operator becomes an additional symmetry operator.
This fact is trivial for all equations
fulfilling (\ref{const1}) 
with $ R=\tau $ due to Proposition 3.1 of \cite{o}. 
In this case however it can be shown that the transformation operator
can be expressed via momentum operator \cite{y}. \\
The general case must be futher studied as two questions arise: \\
- whether the transformation operator is also the symmetry operator 
for a certain class of equations considered above, \\
- can the transformation operator be expressed by deformations
of classical symmetry operators as in the case $ R=\tau $. \\
Let us notice that in the  algebraic 
structure of the example 
studied in \cite{d} the transformation operator is 
not necessary to close 
algebra, however without this operator one can not 
close the co-algebra. \\
We wish to point out some other open questions. \\
In classical theory 
(as well as in discrete models 
\cite{a,b,c}) the consequence of conservation laws
for equations is existence of conserved quantities. 
They were constructed using 
integrals on continuous and discrete space-time
obeying Stokes-type theorem. Once the integral calculus
on quantum space time (\ref{comx})
compatible  with the 
differential calculus (\ref{comp},\ref{com}) will
be developed we shall be able to derive conserved quantities 
for arbitrary linear models. \\
Our aim is also the extension of the presented method
to other types of non-commutative differential calculi, it should be
interesting for braided differentail 
calculus studied by Majid in \cite{l,m,n}.
The promising feature is existence 
of integral calculus which could be applied in 
further construction of integrals of motion \cite{n,z,q}. \\
The other interesting 
problem is systematic study of symmetry operators,
their algebraic and co-algebraic structure for equations
on quantum Minkowski spaces. \\
We hope to come back  to these questions in the subsequent paper.
\section{Acknowledgments}
The author is thankful 
to Professor J. Lukierski and Dr P. Podle\'{s}
for valuable discussions.

\end{document}